\def\ii#1\ff{\textul{#1}}	

\def\exp{\mathrm{exp}}

\def\bbsty#1#2#3{{\bf #1}, (#3) #2}	

\documentclass[psfig,amsmath,amssymb]{cimento_modif}
\usepackage{graphicx}\usepackage{epsfig}
\usepackage{amsmath}\usepackage{amssymb}\usepackage{amsfonts}
\usepackage{bm}
\usepackage{exscale}
\usepackage{dcolumn}    
\usepackage{color}
\usepackage{soul}
\usepackage{ulem}
\title{Collective and dissipative effects in a common microscopic dynamical description}
\shorttitle{Collective and dissipative effects in a dynamical description}
\author{
	H.~Dinh Viet\from{ins:x},
	P.~Napolitani\from{ins:x}}
\instlist{
	\inst{ins:x} Universit\'e Paris-Saclay, CNRS/IN2P3, IJCLab, 91405 Orsay, France}
 
\begin{document}

\maketitle

\begin{abstract}
Depending on the energy regime, the dynamics of heavy-ion collisions reveals a variety of different mechanisms which are attributed to the combination of collective and dissipative effects.
Semi-classical approaches have been successful in describing chaotic regimes at Fermi-energies but they gradually lose precision when extending to collective behaviour and in general when low-energy features become more determinant in the dynamics.
To improve on this description, we propose a theoretical approach starting from the TDHF scheme.
A quantum representation with a moving basis function has been worked out with a double aim.
Firstly, achieving a simplified solution to handle the evolution in time.
Secondly, introducing beyond-mean-field extensions and stochastic contributions. Applications to nuclear collisions at incident energies around low to Fermi energy are presented.
\end{abstract}

%
%
%
%
%
\section{Introduction}

	In heavy-ion collisions there is a wide range of phenomena that is explored depending on various conditions of incident energy, isospin asymmetry, impact parameter and other properties.
	Schematically, the mechanisms could be distinguished into at least three relevant energy regimes.
	First, the low energies up to around 15 MeV per nucleon, where two-body nucleon-nucleon collisions are suppressed due to Pauli blocking in the final states. 
	Therefore, the physics is dictated by the long-range collective effects of the mean-field potential.
	Next from several tens to hundreds of MeV, at the so-called Fermi to intermediate energies, the two-body nucleonic interactions have to be included in addition to the collective behaviour.
	Lastly, the participant-spectator regime at high to relativistic energies, where the short-ranged two-body interaction dominates over the mean-field contribution.
	For each of those energy ranges, there are different dedicated models that are well-adapted to the situation at hand.
	The low energy mechanisms are for example efficiently described within the time-dependent Hartree-Fock (TDHF) framework, whereas the large-amplitude fluctuations at Fermi energies are described in stochastic mean-field approaches (Boltzmann-Langevin equation) \cite{Chomaz2004,Napolitani2013,Colonna2017} or in molecular dynamic models \cite{Ono1992,Feldmeier1990,Aichelin1991}.
	
	However, a challenge for nuclear many-body theories is to properly address the competing types of instabilities at the threshold between Fermi and low energies within a unified picture. 
    At the state-of-the-art, models at Fermi energies tend to lack the mean-field phenomenology to some extent, from isospin drifts to collective modes, and, conversely, models approaching low energy miss large-amplitude fluctuation and lack mechanisms where profound transformation, or even splits, occur in the system.
	Ultimately, a theoretical approach able to cover the transition from Fermi to low energies in one single comprehensive description, would be a well suited framework to study the evolution of fragments and clusters as a function of time and density.
	
	In the following, we are adapting a stochastic TDHF formulation (previous formulations we progress from are \cite{Jouault1998,DeLaMota2020}) by starting from the many-body Schroedinger equation, since this is the underlying physics which determines the collective behaviour of any quantum system at these energies.
	The fundamental point of this approach is the decomposition of the non-local nucleonic wave functions into a set of moving basis functions.
	As a consequence, the mean-field properties are preserved since the nucleonic wave function is not constrained to be localised and furthermore the system is prepared into a set of moving basis functions in order to follow a similar scheme as in the analogous semi-classical Boltzmann-Langevin approach.
	 
\section{Model}

	The motivation of this model is to describe the mechanisms of the Fermi to low energy. 
	Hence, it is important to keep a good description of the mean-field to account for the collective behaviour, while preparing the system for a stochastic collision term as well.
	The mean-field will be given by the one-body density matrix, or equivalently the one-body nucleonic wave functions, which will be determined from a self-consistent TDHF method using a Skyrme-type potential.
	The crucial difference of this model is the decomposition of the nucleonic wave functions $\lvert\varphi_i\rangle$ into a set of moving Gaussian bases $\lvert g_j\rangle$ with varying widths in momentum and configuration space:
	\begin{eqnarray}
	\lvert\varphi_i\rangle &=& \sum_j c_{ij} \lvert g_j\rangle
	\quad \tx{and} \quad 
	\rho = \sum_i \lvert\varphi_i\rangle \langle \varphi_i\rvert 
	\;,\label{eq:decomposition}\\
	\tx{where\ }& &
	g_j(x) = \left( \frac{1}{2\pi\chi_j}\right)^\frac{1}{4}
			 \exp \left[ -\frac{(x-x_j)^2}{4\chi_j} 
			 			 + \tx{i}\frac{\sigma_j}{2\chi_j}(x-x_j)^2
			 			 + \tx{i}k_j (x-x_j)\right]
	\;.\notag	
	\end{eqnarray}	
	The parameters are position $x_j$, momentum $k_j$, configuration width $\chi_j$ and the width correlation parameter $\sigma_j$, ensuring Heisenberg's uncertainty principle.
	It is instructive to mention that in the limit where the decomposition in eq.~\ref{eq:decomposition} is restricted to only one Gaussian, the nucleonic wave function is constrained to a localised, parametrised Gaussian wave packet as it is the case in molecular dynamics.
	By including more Gaussians into the decomposition, the accuracy to represent the non-localised Hartree-Fock wave function is steadily increased, which in turn improves the mean-field description.
	
\begin{figure}[t!]\begin{center}
	\includegraphics[angle=0, width=0.7\textwidth]{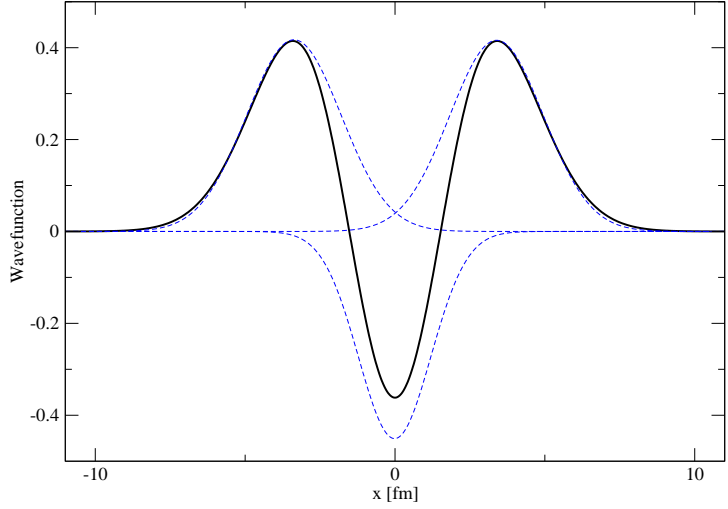}
    \end{center}\caption
{
	Decomposition of an harmonic oscillator state (solid black) into a superposition of three Gaussian base function (dashed blue).
}
    \label{fig:decomposition}
    \end{figure}

	Furthermore, it should be noted that the coefficients $c_{ij}$ of the decomposition are not only positive but also negative, contrary to semi-classical Boltzmann models with test-particles, which have all-equal and positive weights.
	In the present development, the coefficients fit the wave function, at variance with a semi-classical approach, where the coefficients would rather fit the phase-space density.
	By Wigner-transformation, it would be possible to interpret the square of the coefficients to a phase-space weight.
	This is illustrated in fig.~\ref{fig:decomposition} where an initial harmonic oscillator state is decomposed into a sum of three Gaussian basis functions. The central Gaussian needs to have a negative coefficient since the original wave function has a negative extremum. 
	
	In this setup, the framework therefore allows to handle not only a moving set of functions as in semi-classical mean-field approaches but to also correct treatment of the Pauli-blocking. That comes from the fact that the wave functions are not Wigner-transformed, which is why the coefficients in the decomposition will take both positive and negative values. Hence, the Pauli principle can still be checked on the basis of the scalar product rather than the phase-space occupancy. The overlap of two nucleonic wave functions consists therefore of constructive contributions from Gaussians of equal sign but also destructive interferences from Gaussians of opposite sign. 
	
	Using the description of the one-body density $\rho$ in eq.~\ref{eq:decomposition} we derive the evolution equation both for a cooling procedure which follows the gradient of the Hamiltonian and for the mean-field propagation by applying the Euler-Lagrange equations.
	The cooling procedure acts after the change of bases from the Hartree-Fock states to the Gaussian representation.
	This helps to find the suited ground state for the Gaussians as it is done in molecular dynamics.
	Afterwards, the time evolution of the system is fully determined by the equation of motion derived from the Euler-Lagrange equations. 
	
	On top of the mean-field evolution, a description of the collision term in the spirit of the Boltzmann-Langevin One-Body (BLOB) model \cite{Napolitani2017,Napolitani2017_HDR} is adopted in order to take into account large-density fluctuations:
	\begin{eqnarray}
	\tx{i} \hbar \frac{\partial\rho}{\partial t} - \left[ h,\rho \right] &=& I_{UU} + \delta I_{UU}\\
	&=& g \int \frac{\tx{d}^3p}{h^3}
		  \int W(AB \leftrightarrow CD) F(AB \rightarrow CD)\tx{d}\Omega
	\;,\notag	
	\end{eqnarray}	
	where $h$ is the mean-field Hamiltonian, $I_{UU}$ the average Uehling-Uhlenberg collision term and $\delta I_{UU}$ its fluctuating part. In the BLOB approach both terms are combined into a nucleonic collision term where nucleonic degrees of freedom are exploited through a stochastic seed to induce fluctuations in a scheme analogous to the Brownian motion.
	
    In a nucleon-nucleon collision event, the involved nucleonic wave functions are collapsed to a representative Gaussian wave function. The final scattered nucleonic wave packets are then adapted to the final state in such a way that Pauli-blocking is not violated. 
	
\section{Discussion}

    \begin{figure}[t!]\begin{center}
	\includegraphics[angle=0, width=1\textwidth]{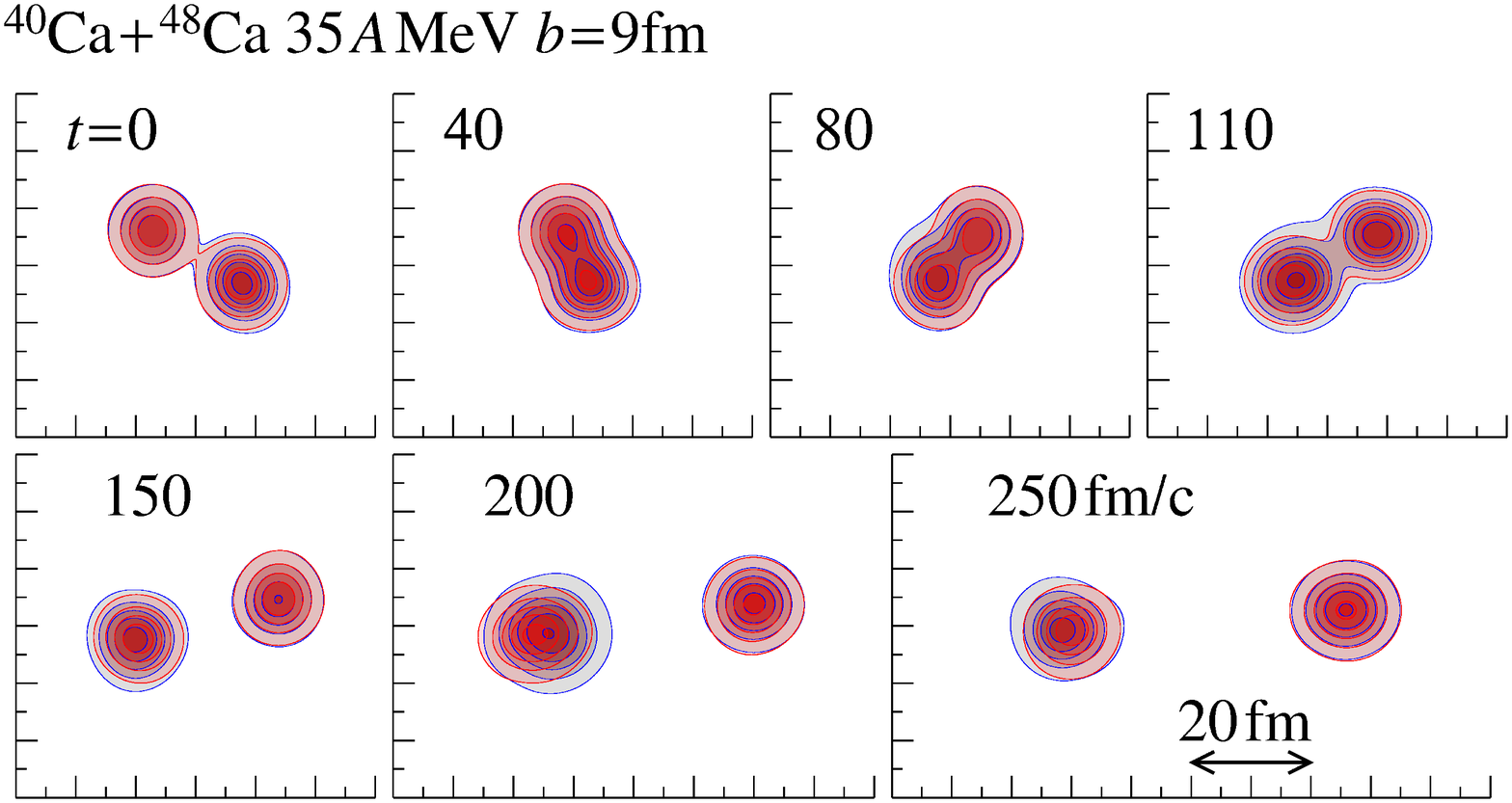}
\end{center}\caption
{
	Snapshots at different times ($t = $ 0, 40, 80, 110, 150, 200 and 250 fm/c) during a peripheral collision of the isospin asymmetric $^{40}\tx{Ca}$ + $^{48}\tx{Ca}$ collision at 35 AMeV with an impact parameter $b = 9$ fm. The red color indicates the proton distribution and the blue color indicates the neutron distribution.
}
\label{fig:Pigmy}
\end{figure}

    In order to test the model, we focus first on the stability of the approach and, later on, we extend to the collision term. 
    As a first example, we chose the asymmetric system $^{40}\tx{Ca}$ + $^{48}\tx{Ca}$ at 35 AMeV (see a recent experimental campaign \cite{CaCaPaper}) and select peripheral collisions with an impact parameter $b = $ 9 fm.
    Under these conditions, we ensure that the collision is well controlled by the mean-field evolution with little to no nucleon-nucleon collisions. 
    In fig.~\ref{fig:Pigmy} this collision is displayed in a contour plot viewed from the side at various times during the collision evolution.
    At $t = $ 0 fm/c, $^{40}\tx{Ca}$ and $^{48}\tx{Ca}$ are impinging from the left and the right, respectively.
    Proton (red) and neutron (blue) density contour plots correspond to initially isometric distributions within the nuclei.
    As the nuclei start to overlap at 40 fm/c, the mean-field effects start to visibly show at 80 fm/c and 110 fm/c. 
    Due to a combination of Coulomb repulsion acting on protons and neutron diffusion effect from the neutron-rich towards the neutron-poor nuclei, we observe in both snapshots at 80 fm/c and 110 fm/c a pull of the neutron distribution of the $^{48}\tx{Ca}$ in the direction of the neck between both nuclei. 
    At the later stages of the evolution, the neck has disappeared.
    However, we can clearly see that the initial drift of the neutron distribution during the contact phase of the collision has induced a dipolar resonance of Pigmy type into the neutron-rich nuclear partner.
    The neutron distribution at 200 fm/c is shifted on the right-side of the neutron-rich nucleus (travelling to the left) whereas at 250 fm/c the neutron distribution has moved to the left side.
    Conversely, we do not see such a pronounced effect in the $^{40}\tx{Ca}$ quasi-nuclei. Consequently, we can deduce that the Coulomb repulsion has less impact than the neutron diffusion \cite{Baran2005} caused by the initial isospin asymmetry during the overlapping phase. 

    Furthermore, another feature worth pointing out is the stability of the final states. In semi-classical transport approaches, the final states do not stay compact over a large amount of time. The reason is that the inherent evolution equation is a Boltzmann equation which leads to a Boltzmann distribution over time. Hence, the distribution tends to smear out as the dynamics goes on. In our case, the final states are very well defined and are not spreading out, which is a direct consequence of the TDHF evolution equations.

    \begin{figure}[t!]\begin{center}
	\includegraphics[angle=0, width=0.8\textwidth]{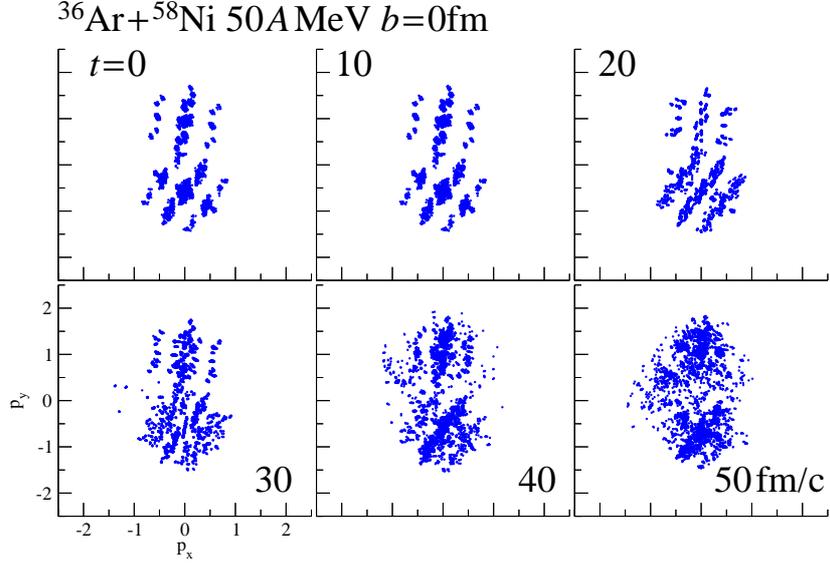}
    \end{center}\caption
{
	Snapshots at different times ($t = $ 0, 10, 20, 30, 40 and 50 fm/c) during a central $^{36}\tx{Ar}$ + $^{58}\tx{Ni}$ collision at 50 AMeV with an impact parameter $b = 0$ fm in momentum space.
}
    \label{fig:PCollision}
    \end{figure}
 
    As a second example, we choose a very dissipative system to test the action of the BLOB-like collision term.
    We consider therefore the central $^{36}\tx{Ar}$ + $^{58}\tx{Ni}$ collision (see experimental campaign \cite{Lautesse}) at 50 AMeV with an impact parameter $b = 0$ fm. In fig.~\ref{fig:PCollision} the evolution of the Gaussian centroids is displayed in momentum space. 
    In the approaching stages from 0 fm/c to 20 fm/c the structural composition of both nuclei remains intact since there are no internal collisions due to Pauli-blocking. 
    As the two nuclei enter their contact phase, progressing from 30 fm/c to 50 fm/c, some portions of the system with the size of a nucleon are pushed outside the Fermi spheres of the respective colliding nuclei. 
    Seemingly to the BLOB approach, each of those portions has been defined by selecting a set of Gaussians which represent a nucleon at the time of a nucleon-nucleon collision. 
    The choice of such sets requires an iterative procedure based on the scalar product and allows to build up the nucleonic degrees of freedom.
    Thus, the scalar product is the equivalent to the phase-space metric in the Boltzmann approaches. 
    On the basis of mean-free-path conditions and the in-medium nucleon-nucleon cross section, the collision can be attempted and further checked for also fulfilling the Pauli principle.
    
\section{Conclusions}
    
    In this proceeding, we presented a new framework which keeps the detail of a quantum mean-field TDHF description, but establishes additionally a foundation to include beyond mean-field extensions of large-amplitude fluctuations. 
    This is achieved by decomposing the nucleonic Hartree-Fock wave function into a set of time-dependent, Gaussian basis functions. 
    We showed the decomposition into Gaussians and the importance of real valued coefficients in order to ensure orthogonality by having both constructive as well as destructive contributions to the overlap between two nucleonic wave packets. 
    The stability of the mean-field evolution have been tested on a peripheral, isospin asymmetric collision, displaying a Pygmy dipolar resonance in the neutron-rich final state. 
    Furthermore, we tested a BLOB-like collision term on a very dissipative mechanism.
    However, the advantage of our framework compared to the stochastic semi-classical transport approach is that multi-particle correlations are naturally introduced by construction, since a set of Gaussians is tracked for each nucleonic wave packet.
    Hence, this new approach should avoid the arbitrariness of introducing nucleonic degrees of freedom from phase-space arguments.
\acknowledgments

\end{document}